\documentclass[12pt,useAMS,usenatbib,usegraphicx]{aastex63}
\usepackage{bm}
\usepackage{xcolor}

\usepackage[left]{lineno}

\definecolor{rkka}{RGB}{219,59,32}

\newcommand{\eb}{\begin{equation}}
\newcommand{\ee}{\end{equation}}

\definecolor{rkka}{RGB}{219,66,32}

\begin{document}
\title{Optical variability of ICRF3 quasars in the Pan-STARRS $3\pi$ survey with functional
principal components analysis}
\author{C.T. Berghea$^{1}$\footnote{E-mail:
ciprian.t.berghea@navy.mil},
V.V. Makarov}
\affil{US Naval Observatory, 3450 Massachusetts Ave NW, Washington DC 20392-5420, USA} 
\author{ K. Quigley} \affil{
Carnegie Mellon University, 5000 Forbes Ave, Pittsburgh, PA 15213-3890, USA}
\author{B. Goldman}
\affil{
Max-Planck-Institut fur Astronomie, K{\"o}nigstuhl 17, D-69117 Heidelberg, Germany}

\date{Accepted . Received ; in original form }

%% \label{firstpage}

\begin{abstract}
We make use of individual (epoch) detection data from
the Pan-STARRS ``$3\pi$" survey for 2863 optical ICRF3 counterparts in the five wavelength bands 
$g$, $r$, $i$, $z$, and $y$, published as part of the Data Release 2. A dedicated method based on the Functional 
Principal Component Analysis
is developed for these sparse and irregularly sampled data. With certain regularization and normalization
constraints, it allows us to obtain uniform and compatible estimates of the variability amplitudes and average
magnitudes between the passbands and objects. We find that the starting assumption of affinity
of the light curves for a given object at different wavelengths is violated for several percent
of the sample. The distributions of root-mean-square variability amplitudes are strongly skewed toward small
values, peaking at $\sim 0.1$ mag with tails stretching to 2 mag. Statistically, the lowest variability
is found for the $r$ band and the largest for the reddest $y$ band. A small "brighter-redder" effect is
present, with amplitudes in $y$ greater than amplitudes in $g$ in 57\% of the sample. The variability
versus redshift dependence shows a strong decline with $z$ toward redshift 3, which we interpret as the time 
dilation of the
dominant time frequencies. The colors of radio-loud ICRF3 quasars are correlated with redshift in a
complicated, wavy pattern governed by the emergence of brightest emission lines within
the five passbands. \end{abstract}


\section{Introduction} 
\label{int.sec}

The crucially important tie of the radio and optical fundamental reference frames relies on
a few thousand International Celestial Reference Frame (ICRF) sources with optical counterparts bright
enough to be detected by the Gaia mission. The optical ICRF sources are not ideal for this
work, with the point-like AGN images being often perturbed by the substrate structures of host or
neighboring galaxies \citep{mak12}. A smaller fraction of reference frame sources resides in double systems
or microlensed images. The intrinsic variability of AGNs is another complication from both an
astrophysical and technical points of view. The astrometric photocenter is determined by the combined flux from the substrate galaxy and the point-like AGN, which may vary in brightness and color. The resulting astrometric perturbations are complex and unpredictable, with magnitudes possibly dependent on redshift and AGN luminosity. Our overarching motivation is to address the emerging problem
of the astrometric link between the two fundamental celestial reference frames, the radio-based ICRF3\footnote{The S/X ICRF3 catalog 
is available for download at \url{http://hpiers.obspm.fr/webiers/newwww/icrf/}.} and the optical Gaia mission frame \citep{pru,bro}.
Using the earlier versions of VLBI-determined and Gaia astrometric catalogs, a significant fraction (a few to several percent) of
reference objects was found to have large radio-optical position offsets \citep{mig16, pet17, mak17}. This can perturb
or completely derail global astrometric solutions that rely on the VLBI positions of a relatively small set of radio loud
AGNs to remove the rigid rotation and other sky-correlated errors \citep{ber16, lin}. This effect was confirmed with the more
accurate Gaia Data Release 2 (DR2) \citep{mak19, pla}. Within the ICRF3 sample, the offset is weakly dependent on redshift,
except for the nearest AGNs, which may in fact be physically offset (dislodged) from the optically dominating centers of their
host galaxies. At greater cosmological distances, relativistic jets dominating the radio photocenters may be shifted from the
central engines by measurable amounts. Some of these interpretations can be indirectly tested using light curves and
general variability estimates. {\sc We would ultimately like to know if the objects with significant position offsets also
display greater optical variability.}
In this paper, we are looking for answers to these questions: 
\begin{itemize}
\item how variable are the ICRF3 quasars overall? 
\item how do the amplitudes of variability compare between the five passbands of Pan-STARRS?
\item does the degree of variability statistically
depend on redshift?
\item what are the typical colors of ICRF3 quasars and are they dependent on redshift?
\item do ICRF3 quasars
appreciably change colors due to brightness variations?
\end{itemize}

Quasars are intrinsically variable sources in the optical continuum, which helps to distinguish them from stars in large
photometric surveys \citep{vanden, haw83, van}. In a large magnitude limited sample, stars always have statistically smaller
``amplitudes" of variability. The character of photometric variability of stars is also different from that of quasars
and AGNs. While the stellar light curves often have periodic or quasi-periodic character caused by internal pulsations
or binary effects, or exhibit randomly placed short-term flares, quasars' flux appear to follow a stochastic, unpredictable
pattern, which is mathematically closer to a random process. These properties make the definitions of both ``variability amplitude"
and ``mean magnitude" model-dependent, to be inferred from
fitted parameters of a chosen stochastic model. Many studies described the degree of variability in terms of empirically
determined structure function ${\rm SF}(\Delta t)$, where $\Delta t$ is the time difference between two measurements.
The square of first-order structure function within the simplest random process models can be interpreted as the sample estimate of the variance of the magnitude difference
between two measurements separated by $\Delta t$ in time \citep[a more general definition of structure function can be found in, e.g.,][]{kas15}. It can be presented as
\eb
{\rm SF}^2(\Delta t) = 2\sigma_m^2\, (1-\rho(\Delta t)) + 2\sigma_n^2,
\ee
where $\sigma_m^2$ is the variance of the true magnitude, $\sigma_n^2$ is the variance of observational noise, and
$\rho(\Delta t)$ is the autocorrelation function, {\it assuming} that 1) the true magnitudes and noise are statistically
independent; 2) the expectancy of noise is zero (no bias); 3) the noise is uncorrelated (no systematic errors); 4)
the process is stationary, i.e., the variance of true magnitude is independent of the true magnitude. It is common
to further assume that the autocorrelation function equals 1 at zero time lag and smoothly declines to zero
at a certain characteristic time lag $\tau$. Somewhat arbitrarily, this expectation leads to common attempts to
model the structure function as a power law with a certain slope $\gamma$ and characteristic lag $\tau$, to be
determined from observation. Certain models are mathematically consistent with this expected behavior of the
autocorrelation function, most notably, the CARMA$(1,0)$ continuous random process, which motivated its
application in AGN light curve analysis for interpolation and forecasting \citep[e.g.,][]{kel14, mor}. These models
are empirical rather than physical, as the physics behind quasar variability is likely to be complex and multivariate
in nature. CARMA$(p,q)$ models with greater number of free model parameters may be closer to the physical reality \citep{Kasliwal+2017}. Even within its limited scope and implicit assumption, the three parameters in the above equation
are highly dispersed among individual objects in a sample. To gain some confidence in the derived models, large
samples of data are often utilized, such as spectroscopically confirmed quasars from the SDSS \citep{van}.
The result may be distorted by trying to represent a diverse population of light curves with a single and simple fitting function
and a set of parameters. Alternatively, the structure function and random process models can be meaningfully estimated for
individual objects given a very dense and continuous observational cadence, such as a Kepler mission light curve
\citep{weh}, but this severely limits the available sample.

In this paper, we attempt to find a compromise solution and to gain some insight into the optical properties of individual
ICRF3 objects while making as few model assumptions as possible.  The method of data analysis should be robust enough to process
the sparse and non-uniform data sets for each individual object, rather than treat the data as a single statistical ensemble.
We therefore develop a specialized Functional Principal Component Analysis (FPCA) approach, which is flexible and amenable to
additional regularization constraints and Bayesian prior. The use of FPCA in astronomy is gaining momentum \citep[e.g.,][]{he},
but this is the first application of this technique to survey-type AGN light curves, to our knowledge.

\section{The sample}
\label{samp.sec}
We analyze astrometric and photometric epoch data collected by the Pan-STARRS DR2 (PS1-DR2) for a sample of 2863 ICRF3 radio sources. PS1 is a wide-field imaging system, with a 1.8~m telescope and 7.7~deg$^2$ field of view, located on the summit of Haleakala in the Hawaiian island of Maui. The 1.4 Gpixel camera consists of 60 CCDs with pixel size of 0.256 arcsec \citep{ona08, ton09}. It uses five filters (g$_{P1}$, r$_{P1}$, i$_{P1}$, z$_{P1}$, y$_{P1}$, hereafter $grizy$), similar to the ones used by the Sloan Digital Sky Survey \citep[SDSS;][]{york00}.The Pan-STARRS photometric system is described in
\citet{ton}.  The largest survey PS1 performed was the 3$\pi$ survey, covering the entire sky north of $-30\deg$ declination. The epoch data for the 3$\pi$ survey used in this paper was made available to the public in Jan 2019 as part of the second data release (DR2).

We started by cross-matching all 4536 ICRF3 sources with the MeanObjectView PS1 catalog in the public archive at the  Space Telescope Science Institute (STScI)\footnote{https://outerspace.stsci.edu/display/PANSTARRS/Pan-STARRS1+data+archive+home+page}, 
using a $3\arcsec$ cone search. This resulted in 3348 objects with at least one match, 133 of which had more than one counterpart. 
Color images were extracted for all these objects using the cutout server at STScI\footnote{http://ps1images.stsci.edu/cgi-bin/ps1cutouts}.
We removed 194 sources which had a significant offset $0.2\arcsec$ between their PS1 and ICRF3 positions. 
We also removed sources which had nearby optical companions (within $\approx 0.5\arcsec$) because they could affect the photometry.
We use the MeanObjectView filter flag data and visual inspection to remove all sources that showed extended structure.
This process left 2896 sources that were deemed acceptable for this study.

The epoch data were extracted from the archive for all these sources (from the view Detection in the PS1 archive). 
We finally removed 14 sources with fewer than 10 observations,
while 19 were missing in the epoch data. These are all inside a specific PS1 observation stripe (dec range 54.2 to 57.6 degrees) due to errors in processing of PS1 DR2 data (Rick White, private communication). The final sample contains 2863 ICRF3 sources.
The epoch data has been cleaned for outliers using a Gaussian kernel with probability density $<$0.05, 
described in detail elsewhere \citep{ber16}. The purpose of this cleaning was to remove rather frequent outliers generated by flukes and real double
or multiple sources within the search radius of $3\arcsec$. These outliers have usually discordant magnitude
estimates and impact the light curve analysis, in some cases resulting in complete failure.

The epoch data for each contains the IVS name of the associated radio source, ICRF3 coordinates (right ascension and declination at J2000), PS1 measurements of $g_{\rm PS1}$, $r_{\rm PS1}$, $i_{\rm PS1}$, $z_{\rm PS1}$, 
$y_{\rm PS1}$ magnitudes, individual astrometric determinations (epoch RA and Dec), associated formal errors,
and other observations and metadata. Not every object has observations in each of the 5 filters. The typical
number of detections per object is 60, so that there are on average only 12 observations per filter, but
the numbers are very non-uniformly distributed. The time span of the survey is about 4 years. The observational
cadences are sparse and non-uniform.

\section{The character of light curves}
Active galactic nuclei (AGN) are intrinsically variable sources. The light curves are not fully
self-correlated or periodic, and are believed to be better represented by stochastic,
memory-less continuous processes, such as damped random walk or autoregressive models \citep{and}. 
The spectral power density of variability is {\bf often} assumed to approximately scale
with frequency as $f^{-2}$ \citep{col}. More recent evidence based on well-sampled Kepler mission and OGLE data suggests significantly steeper scaling dependencies, turning points, and intrinsic scatter between individual objects of the same class \citep{mus, zu, kas15, smi}. The characteristic time scale of variations is about 1 month,
but significant variations have been detected even on the intra-night time scales. PS1 observations span
about 4 years, but the observing cadence was irregular, with batches of epochs separated by long
spans with no data at all (Fig. \ref{lc.fig}). The degree of variability appears to be different among the sample sources too.
To complicate the analysis even more, PS1 magnitudes were obtained in 5 different filters. The distribution
of epochs is systematically different for each of the Pan-STARRS filters. A given object was never
observed with two filters on the same night. Brightness variations of quasars are not always achromatic
\citep{haw}, for a few physical reasons discussed below. Thus, we can not generally assume that magnitudes follow
the same light curve in each filter with only a constant shift between them.

Let $L(t)$ be a light curve of a given QSO as a function of time expressed in magnitudes. We assume that
$L(t)$ is a continuous and smooth (differentiable) function. This function is often modeled using a
continuous autoregressive (CAR) process for survey-type light curves \citep[e.g.,][]{kel}, further developed into the more general and
accommodating CARMA$(p,q)$ model \citep{kel14,sim,Kasliwal+2017}.
In our case, there are too few data points in each of the five filters to use this method. The main
results of the CAR-fitting are the characteristic time scales and the level of random noise component,
which have statistical bearing on certain physical parameters of interest, e.g., the $L/L_{\rm Edd}$
ratio. Given the sparsity of the data, 
we set more modest goals for this study: estimate the degree of variability in a consistent
and reproducible way for each ICRF objects in each filter, as well as ``average" magnitudes and possible
chromatic deviations in the light curves.

\section{Functional PCA}
The basic idea of the FPCA approach is that the underlying light curve can be represented by a finite
set of basis functions, and the observed discrete series of magnitudes can be well represented by
a small number of orthogonal discrete functions called Principal Components. Each of the PC functions
is completely represented by its projection onto the set of basis functions. Here we use the Fourier
functions of time augmented with a linear trend as the basis in the space of continuous functions:

\eb
L(t)=\sum_{i=1}^{K} C_i \cos\left(\frac{2\pi\,i\, (t-t_0)}{\Delta t}\right)+ \sum_{i=1}^{K} S_i \sin\left(\frac{2\pi\,i\, (t-t_0)}{\Delta t}\right)+
S_0\,\frac{t-t_0}{\Delta t},
\label{l.eq}
\ee
where $t_0$ is the mid-epoch of observations, $\Delta t$ is the time span of observations. The presence
of the linear trend term is justified by the limited time span of PS1 light curves (roughly, 4 years)
and the likely possibility that the longer-term light curve variation will not be captured by the
periodic Fourier terms. The limiting index $K$ defines the highest frequency of the fitting functions.
Technically, any discrete set of magnitudes can be {\it exactly} represented by this decomposition
if $K$ corresponds to the Nyquist frequency of the {\it smallest} time interval of consecutive observations,
$\delta t_{\rm min}={\rm min}(t_{i+1}-t_i)$, i.e., $K={\rm Ceil}(\Delta t/(2\,\delta t_{\rm min}))$. In
practice, we found it practical to restrict the model to fitting only variations on time scales
equal to or longer than 2 weeks (14 days). If true variations within 2 weeks are present in the data,
they are not captured by the fitting model, thus raising the residual errors. This is first in a series
of sacrifices we have to make to obtain a stably solvable problem.

We note that Eq. \ref{l.eq} is already a step away from the idea of using basis functions, because while
the Fourier terms are mutually orthogonal, the linear trend term is generally not. This has no practical
consequence for the following processing, because the discretized functions will be orthogonalized
in the following PCA.

Let us re-write Eq. \ref{l.eq} as
\eb
L(t)=\sum_{i=0}^{2K+1} \alpha_i Y_i(t),\label{alpha.eq}
\ee
where $Y_i(t)$ are the Fourier terms and the linear trend, in arbitrary order. The measurements
are taken at a set of discrete epochs $t_j$, $j=1,\ldots, n$. The functions $Y_i(t)$ can
be discretized at the same cadence, $Y_{ij}=Y_i(t_j)$, yielding a set of linear equations. These
equations can not be solved for the coefficients $\alpha_i$ because the system is severely
indeterminate. In order to further limit the number of unknowns, we perform PCA of the discretized
functions $Y_{ij}$. 

Let $\bm{Y}=(\bm{Y}_1,\bm{Y}_2,\ldots,\bm{Y}_{2K+1})$ be the matrix composed of column vectors
$\bm{Y}_i$, which include values of the fitting function $Y_i$ at times $t_j$, $j=1,\ldots, n$. The size
of the matrix is $n\times (2K+1)$. If
\eb
\bm{U}\,\bm{\Sigma}\,\bm{V}^T=\bm{Y} \label{svd.eq}
\ee
is the singular value decomposition (SVD) of the matrix, the diagonal of $\bm{\Sigma}$ includes the
singular values $\sigma_j$ in descending order, $\sigma_1\ge\sigma_2\ge \ldots \ge\sigma_n$, and the leading
$n$ columns of $\bm{U}$ are the singular (basis) vectors spanning the space of vectors $\bm{Y}_i$.
This decomposition allows us to not only decrease the number of fitting functions to the number of
observations, but to further restrict the solution to a smaller number of the most significant functions.
The first column of $\bm{U}$ is the most significant vector, which is most representative of the actual
observational cadence, called the first principal component. The exact representation of the discretized
functions $\bm{Y}_i$ is achieved with $n$ principal components, but it turns out that a sufficiently
close representation is obtained with a smaller number of components corresponding to the largest $\sigma_j$.
We chose to make a cut at $j=n_\sigma$ such that $\sigma_j\ge0.1\,\sigma_1$ for $j=1,2,\ldots,n_\sigma$,
and $\sigma_j<0.1\,\sigma_1$ for $j=n_\sigma,\ldots,n$, i.e., use only the components with corresponding
singular values greater or equal to one tenth of the largest singular value. 

The emerging problem can now be written as
\eb
L(t)=\sum_{i=1}^{n_\sigma} \beta_i U_i(t),\label{beta.eq}
\ee
where the continuous functions $U_i$ are certain linear combinations of the much greater number of functions $Y_i$
defined by the SVD in Eq. \ref{svd.eq}. When discretized on a particular cadence of times of observation,
the overdetermined system of linear equations
\eb
\bm{L} = \sum_{i=1}^{n_\sigma} \beta_i \bm{U}_i,\label{beta0.eq}
\ee
can be solved by the Least Squares method, but, using the orthonormality of PCs, the straightforward solution
is obtained by projection
\eb
\beta_i=\bm{U}^T_i\,\bm{L}. \label{sol.eq}
\ee
Now recall that the vector $\bm{L}$ includes all observations of a given object in all the 5 bands, and
it describes only the variation of brightness around a constant magnitude. The constant magnitude and the
amplitude of variation can be different for different filters. The next tier of this model is to introduce
an affine transformation from the basic light curve $\bm{L}$ to a filter-specific light curve $\bm{L}^{(f)}$:

\eb
\bm{L}^{(f)}=m^{(f)}+a^{(f)}\,\bm{L_f}, \label{gamma.eq}
\ee

where the vector $\bm{L}^{(f)}$ includes all the observed magnitudes for a specific filter, and $\bm{L_f}$
is the basic variation function $L(t)$ computed at the times of observation in that filter. $\bm{L}^{(f)}$ is the filter-specific observed light curve, while $\bm{L}_f$ is the corresponding portion of the fitted light curve $\bm L$ in Eq. \ref{beta0.eq}. The fitting coefficients
$m^{(f)}$ and $a^{(f)}$ (up to 10 for each object) are the main products of this technique. Equations (\ref{beta.eq})
and (\ref{gamma.eq}), each being perfectly linear, have to be solved in combination, bringing up a nonlinear
problem. Also observe that this problem is degenerate, because the coefficients $\beta_i$ can be multiplied
by a number and $a^{(f)}$ divided by the same number producing the same fit. To resolve this degeneracy, and
also to be able to compare the variability amplitudes between the objects, a standard normalization of the
basic variability vector $\bm{L}$ is required. The natural choice is to set the norm of the vector to its rank:
\eb
|| \bm{L}|| = \bm{L}^T\,\bm{L}=\sum_{i=1}^{n_\sigma} \beta_i^2 = n_\sigma.
\ee
This choice of normalization allows us to interpret the resulting $m^{(f)}$ as the estimated
``long-term average" magnitude and $a^{(f)}$ as the estimated ``characteristic amplitude" of variability in the
same units. 

The fitting procedure for each quasar begins with evaluating the original fitting functions $Y_i$ at times of observations $t_j$. The resulting $n$ by $2K+1$ matrix provides the principal components $\bm{U}_i$ per Eq. \ref{svd.eq}, which define the functional principal components $U_i(t)$. The observed magnitudes are not used at this step, which should be done only once. The subsequent optimization problem is solved iteratively, starting with initial estimates of $m^{(f)}$ and $a^{(f)}$ for each of the five filters. We use the median magnitudes and MAD values as initial estimates.
Eq. \ref{gamma.eq} is solved in inverse in each iteration to obtain the segments $\bm{L}_f$, which are further
combined into $\bm{L}$ at the right cadence and the coefficients $\beta_i$ are computed per Eq. \ref{sol.eq}. To preserve the normalization of these coefficients, a quantity called scale is computed
\eb
\label{G.eq}
\Gamma=n_\sigma/\sum_{i=1}^{n_\sigma} \beta_i^2,
\ee
and applied, after which the forward loop of iteration begins with computing $\bm{L}(f)$ and solving Eq. \ref{gamma.eq} by linear regression. The convergence of this method is normally fast, reflected in the scale
numbers tending to unity and the updates to $m^{(f)}$ and $a^{(f)}$ tending to zero. 

\section{Regularization}
One of the products of this FPCA technique is the continuous function $L(t)$ representing the fitted underlying
magnitude variation of the object. It can be easily reconstructed from the final set of $\beta_i$, basis functions
$Y_i(t)$, and matrices $\bm{\Sigma}$ and $\bm{V}$. Due to the large number of basis functions involved in the fit,
covering the entire range of observable frequencies, the estimated light curves can take large, unrealistic
deviations in the gaps between observations, where the solution is not constrained by the data. To regularize
the solution better, and to introduce a Bayesian prior into the fitting process, we chose to weight the
set of basis Fourier functions by
\eb
\hat{Y}_i(t)=Y_i(t)/\sqrt{i}.
\label{wh.eq}
\ee
This helps to obtain smoother and better behaved light curve solutions and to introduce the previously
found inference that the spectral power density is a declining function with frequency (although an exact
dependence is problematic for the wide range of timescales from one week to several weeks involved in
our case). Even with this regularization, the fits tend to show unreasonable-looking wiggles between 
widely separated nodes. This is the direct consequence of the long gaps in the observing cadence of Pan-STARRS due to observing seasons for any ground-based telescope.
Unfortunately, the fits can not be used to interpolate the light curves between the data points.

The fitted light curve $L(t)$ is not used in this study to interpolate or predict the magnitudes between the points of observation, so the utility of this additional regularization may be questionable. However, experimenting with the data in the general sample, we found a small fraction of objects with one or two apparent outliers in one of the bands, which show a sudden change against a general trend confirmed by nearly simultaneous measurements in other bands. The available formal errors and quality flags do not allow us to objectively distinguish such occurrences, while a single deviant point may result in a poorly justified high-frequency component of the fitted function and bias the amplitude estimation in the unaffected bands. Therefore, this regularization helps to slightly reduce the rate of perturbed solutions. As a downside, the fitted light curves cannot be used to investigate the spectral power distribution of ICRF quasar variability.

\begin{figure*}
\includegraphics[width=150mm]{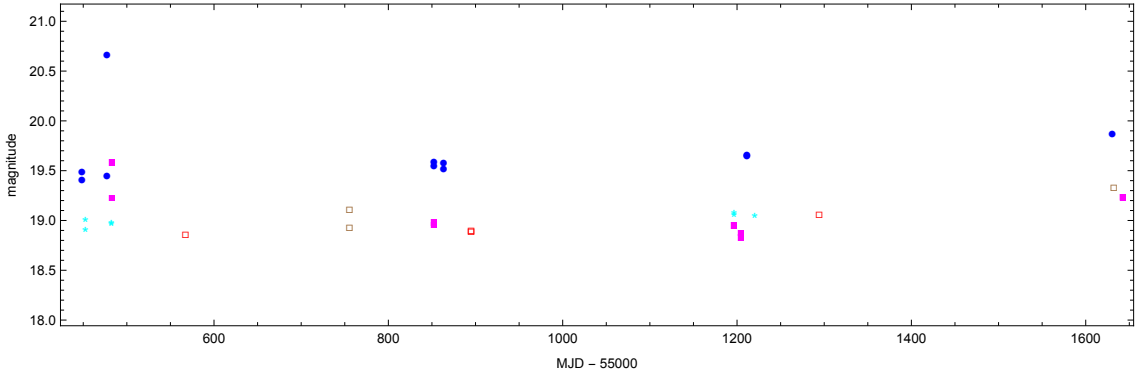}
\caption{A Pan-STARRS light curve for one of the fundamental reference frame quasars. Measurements with
different filters are shown with different symbols: blue circles for $g$, cyan circles for $r$, filled magenta
squares for $i$, open red squares for $z$, and open brown squares for $y$. Most of the formal magnitude errors are between 0.02 and 0.05 with the exception of $y$-filter measuremnts where half of the errors are between 0.10 and 0.014 mag. Note the large gaps between observations due to observing seasons, possibly one faint outlier at the beginning of the survey, and the variable $g-i$ color of the
object. Also note that most observations are not simultaneous in different filters}
\label{lc.fig}
\end{figure*}

Propagation of random measurement errors into the estimated parameter space is an important and nontrivial aspect of light curve analysis. Our method has been specifically designed to work for the extremely sparse and irregular observational cadences of the Pan-STARRS general survey obtained with five different filters. These five cadences are combined into a single light curve allowing free affine mapping between them. The underlying signal is assumed to be homomorphous, but each individual measurement is perturbed by a random uncorrelated error, which violates this assumption. As a result, a perfectly constant source processed with the FPCA method will show small RMS variability amplitudes that are merely random coherences between the filter-specific data samples. It is a well-known fact that large gaps in light curves can give rise to magnified random errors in the estimated variability amplitudes, for example, in the traditional periodogram analysis. This happens because the weak constraint of condition equations results in a wider dynamical range of eigenvalues and emergence of low-significance principal components. The advantage of our method is that this harmful build-up of random error propagation via the least significant principal components with small eigenvalues is avoided by limiting the fit to only most significant components, which are relatively well constrained. Still, the power spectrum density (PSD) of uncorrelated noise is flat, and the estimated amplitudes are affected by its presence in the data. The additional regularization of the solution by weighting of initial Fourier terms (Eq. 11) and the transformation from Fourier functions to functional principal components makes it difficult to analytically estimate the remaining propagation. 

The best approach is to use Monte Carlo simulations perturbing each data point by random numbers with the specified formal variances and zero mean and repeating the FPCA fitting and estimation multiple times. Even non-Gaussian distributions of measurement error can be tested and experimented with, but, unfortunately, Monte Carlo simulations are quite computationally costly. We performed limited experiments with a few selected objects. For the ICRF3 quasar IVS $0000-197$, which is listed in Table~1, we performed Monte Carlo trials perturbing the initial data 300 times and recomputing the mean magnitudes “cv”and RMS amplitudes “av” (the FPCA part does not have to be recomputed because it is defined by the cadence only). The formal Pan-STARRS errors for this object range from 0.025 to 0.199 mag, the greater values being mostly associated with the blue {\it g} and {\it r} passbands where this red quasar is faint. The errors are comparable to the estimated parameters, and, not surprisingly, affect the results. Using $1.5\times$MAD as a robust statistic proxy for standard deviation with the emerging non-Gaussian sample distributions, we obtain these estimates of the amplitude in the {\it grizy} bands: 0.11, 0.09, 0.22, 0.20, 0.23 mag. The blue passbands amplitudes seem to be less dispersed, however, their sample distributions are steeper on the short side and have long tails stretching to values above 1 mag. The median amplitudes over 300 trials are 0.56, 0.54, 0.75, 0.52, and 0.52 mag, respectively. These results are not too far from the single-point estimates in Table~1, except for the {\it g} filter, where observation errors conspired to produce a result at the high end of the sample distribution. 

AGN light curve methods based on random process models also suffer from measurement errors in the input data with large gaps and irregular cadence. These techniques are focused on the PSD parameter rather than the RMS amplitude, because the former is believed to be related to the underlying physical processes in the accretion disks. In the elaborate CARMA$(p,q)$ method of posterior random process fitting, for example, several free parameters are directly related to the autocorrelation function and the relative strength of the Lorentzian PSD components \citep{kel14}. The post-fit standardized residuals are used to select the degrees of the model striving to achieve a Gaussian-looking distribution. This is suitable for better sampled and monochromatic light curves, but may not work for the Pan-STARRS data. More flexible and generalized CARMA-based models have been proposed using the Green’s functions to infer the random process differential equation \citep{Kasliwal+2017} and the physical factors behind it, but they also require dense and uniform observational samples.

\section{Results}

In this study, we only processed 2714 objects with at least 2 observations in each filter. The remaining
433 objects were either too faint or too poorly observed. 
In addition, a significant number of observations are faint and have therefore large errors. To identify these we calculated the signal-to-noise ratio (SNR) as the FPCA-fitted amplitude of variability divided by the median error and we find that 25\% of all observations have SNR~$<$~3. We decided not to remove these and reduce the sample as we found that the basic results are not changed, but we provide the SNR for each observation.

The main product of the FPCA analysis is a
table with the following determined parameters for each object and each filter: the nominal magnitude
$m^{(f)}$, the variability amplitude $a^{(f)}$, the scale parameter $\Gamma$ (see Eqs. \ref{gamma.eq} and
\ref{G.eq}), as well as the median observed magnitude, the observed median absolute deviation (MAD) and the  signal-to-noise ratio as described above.
These data are collected in Table~\ref{res.tab}, which is also available online in full form. 

 \begin{table}
 \centering
 \caption{Catalog of ICRF3 variability parameters.}
 \label{res.tab}
 \begin{tabular}{cccccccccc}
 \hline
  IVS ObjID     &  N   & Scale & MAD Resid & Band & cv & av & Median & 3 $\times$ MAD & SNR \\
  	& 	& 	& mag &  & mag & mag & mag & mag & \\
 1 & 2 & 3 & 4 & 5 & 6 & 7 & 8 & 9 & 10 \\
 \hline
 \text{0000-160} & 49 & 0.98 & 0.067 & \text{r} & 21.215 & 0.104 & 21.248 & 0.169 & 0.8 \\
 \text{0000-160} & 49 & 0.98 & 0.067 & \text{i} & 20.549 & 0.243 & 20.552 & 0.345 & 2.8 \\
 \text{0000-160} & 49 & 0.98 & 0.067 & \text{z} & 20.348 & 0.193 & 20.348 & 0.312 & 1.6 \\
 \text{0000-160} & 49 & 0.98 & 0.067 & \text{y} & 19.779 & 0.326 & 19.779 & 0.526 & 1.8 \\
 \text{0000-197} & 54 & 0.99 & 0.037 & \text{g} & 20.006 & 1.121 & 19.985 & 1.256 & 32.6 \\
 \text{0000-197} & 54 & 0.99 & 0.037 & \text{r} & 19.782 & 0.472 & 19.727 & 0.633 & 13.3 \\
 \text{0000-197} & 54 & 0.99 & 0.037 & \text{i} & 19.469 & 0.421 & 19.484 & 0.394 & 9.6 \\
 \text{0000-197} & 54 & 0.99 & 0.037 & \text{z} & 19.070 & 0.599 & 19.067 & 0.660 & 13.0 \\
 \text{0000-197} & 54 & 0.99 & 0.037 & \text{y} & 18.876 & 0.449 & 18.876 & 0.475 & 4.6 \\
 \text{0000-199} & 63 & 0.91 & 0.018 & \text{g} & 19.168 & 0.052 & 19.187 & 0.110 & 2.3 \\
 \text{0000-199} & 63 & 0.91 & 0.018 & \text{r} & 19.149 & 0.029 & 19.142 & 0.063 & 1.2 \\
  \hline
 \end{tabular}
 \tablecomments{Columns description: 1) IVS object name; 2) total number of observations used in all five filters;
 3) scale $\Gamma$; 4) post-fit residual, MAD; 5) PS1 band; 6) FPCA-fitted mean magnitude; 7) FPCA-fitted
 amplitude of variability; 8) observed median magnitude; 9) observed dispersion, MAD times 3; 10) signal-to-noise ratio.}
 \end{table}

\begin{figure}
\includegraphics[width=85mm]{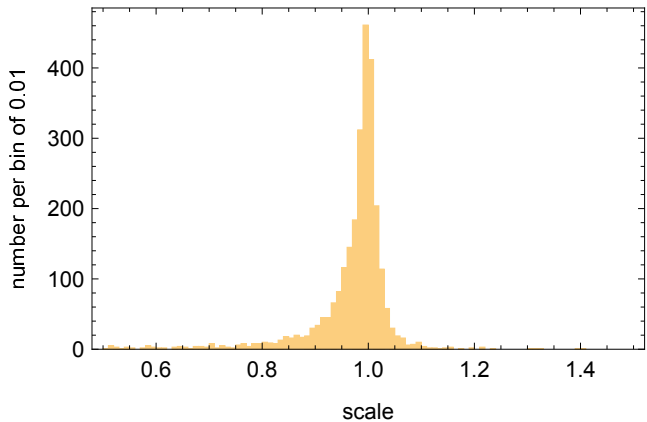}
\caption{Distribution of derived scale parameters $\Gamma$.}
\label{scale.fig}
\end{figure}

Fig. \ref{scale.fig} shows the histogram of the scale parameter $\Gamma$, one per object. There is a sharp and well
defined core with a modal value close to unity, as expected, but a considerable fraction of objects deviate from
this value, especially on the negative side. The tails of this distribution indicate that not all of the ICRF
sources obey the main assumption of our model, that the light curves in different filters differ only by constant offsets and constant multipliers (see Eq. \ref{gamma.eq}). Indeed, the example shown in Fig. \ref{lc.fig} illustrates an
obvious departure from this assumption with the $g-i$ color changing from approximately 0 at the beginning of
the cadence to about $+0.6$ mag. Such perturbations result is poor or absent convergence of the estimated scale value.
The comparability of the amplitudes is also degraded, as the model is not capable of capturing large color variations.

The total number of objects with $|\Gamma-1|>0.5$ is 63. Some of these objects are simply too faint, such as 
IVS B$2143-236$, which was not detected in $g$ and $y$ and just barely detected in $r$. 
Some are double or lensed sources, e.g., IVS B$2057+235$, which has a reddish companion at 1.3 arcsec separation. 
A scale parameter much deviating from unity is a
warning sign for problem objects that may not be suitable as radio-optical reference frame sources. They seem to include
more than one optically luminous component with drastically different light curves, mostly at the blue end of the range.

\begin{figure*}
\includegraphics[width=55mm]{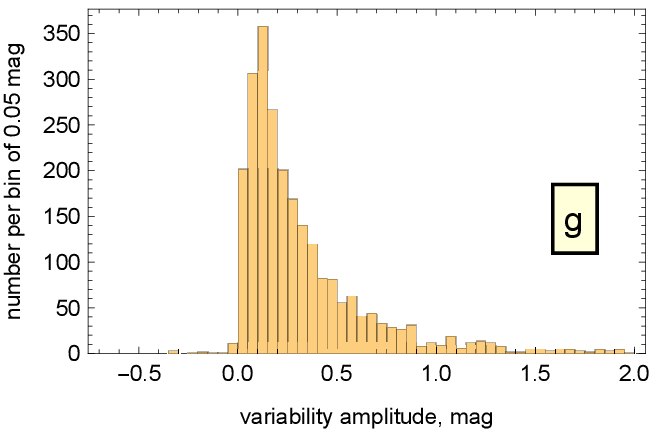}
\includegraphics[width=55mm]{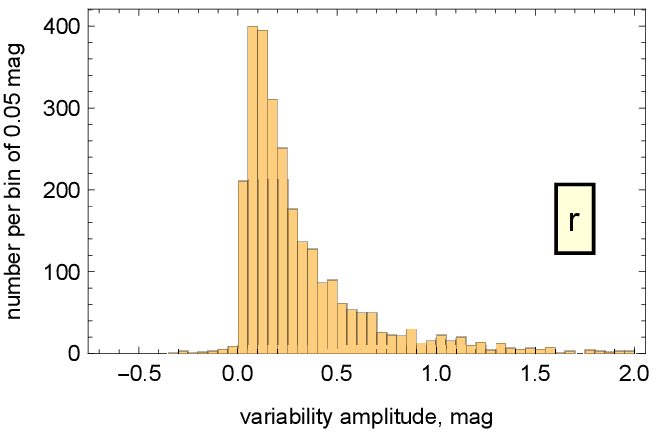}
\includegraphics[width=55mm]{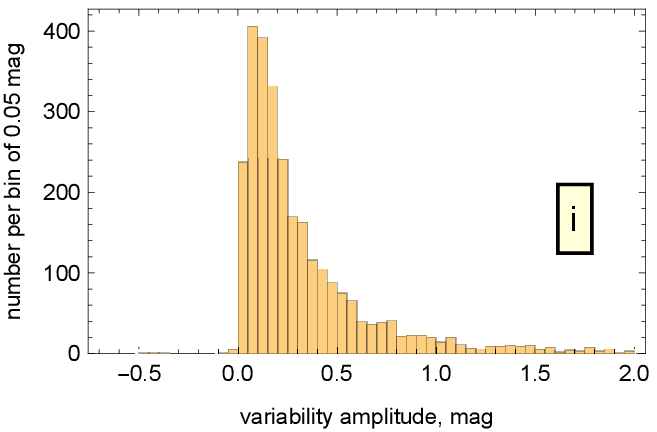}
\includegraphics[width=55mm]{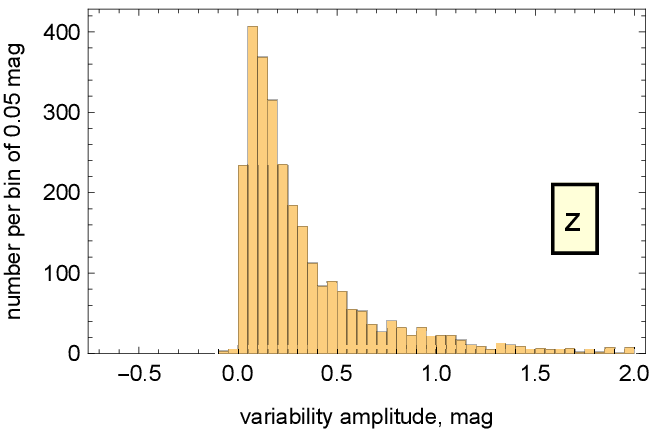}
\includegraphics[width=55mm]{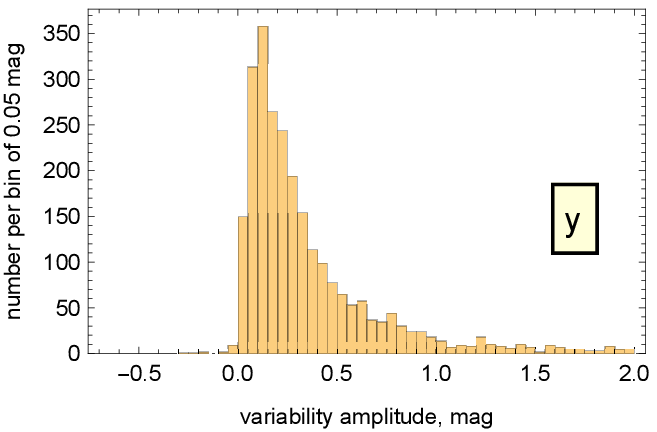}
\caption{Histogram of variability amplitudes $a^{(f)}$ of ICRF3 quasars in all five bands of Pan-STARRS.}
\label{ai.fig}
\end{figure*}

Fig. \ref{ai.fig} shows the histograms of the variability amplitudes $a^{(f)}$ (in the Pan-STARRS bands
$g_{\rm PS1}$, $r_{\rm PS1}$, $i_{\rm PS1}$, $z_{\rm PS1}$, and $y_{\rm PS1}$). The distribution is
sharp-peaked at approximately 0.1 mag, with a slower drop extending to a shallow tail up to 2 mag and larger. 
We quantify the
distributions in three different ways. Firstly, we determine the median of each distribution,
which signifies the ``typical" amplitude. We also fit a Rayleigh distribution
with a free scale parameter $\sigma$ into each of the five histograms. Finally, we determine
the most frequent amplitude, which is called the commonest, at a resolution of one hundredth.
The results are presented in Table~\ref{h.tab}. The dependence of variability amplitude on
wavelength is concave, with the largest values observed for the reddest bands $z$ and $y$, and
the smallest for mid-range bands $r$ and $i$. The peculiar shape of the amplitude distributions is
reflected in the commonest values being much smaller than the median, which are, in their turn,
systematically smaller than the scales of the Rayleigh fits. All quasars are intrinsically variable
sources, however, the modal and most frequent values are rather small.
The larger degree of variability at the reddest wavelengths is not what one would expect from an
intuitive assumption that the contribution from the host galaxy, which is constant, is greater for the red bands.
The method of cumulants that we used to estimate the distribution scale is sensitive to the significant departures
from the assumed Rayleigh PDF, especially, to the extent of the tail. But the more robust median estimates 
confirms that there is
a tendency for the ICRF objects to vary more in the reddest bands $z$ and $y$. This result is also supported
by comparison of the amplitudes for the same object. 57\% of the sample have $a^{(y)}>a^{(g)}$, and
53\% have $a^{(z)}>a^{(r)}$, while $a^{(i)}$ is statistically the smallest. On the other hand, Pan-STARRS observations are systematically noisier in the reddest $y$ band, and a better error budget analysis is in order to confirm this unexpected outcome.

 \begin{table}
 \centering
 \caption{Variability amplitude distribution parameters for Pan-STARRS bands:
 the median value, the scale $\nu$ of the best-fitting Rayleigh distribution, and the commonest value.}
 \label{h.tab}
 \begin{tabular}{@{}lrrr@{}}
 \hline
  Band     &  Median   & Rayleigh $\nu$ & Commonest\\
  			& mag	& mag 	& mag\\
 \hline
 $g_{\rm PS1}$ & 0.220 & 0.302 & 0.13\\
 $r_{\rm PS1}$ & 0.208 & 0.284 & 0.09\\
 $i_{\rm PS1}$ & 0.209 & 0.308 & 0.07\\
 $z_{\rm PS1}$ & 0.217 & 0.319 & 0.07\\
 $y_{\rm PS1}$ & 0.238 & 0.345 & 0.11\\
  \hline
 \end{tabular}
 \end{table}

\begin{figure*}
\includegraphics[width=50mm]{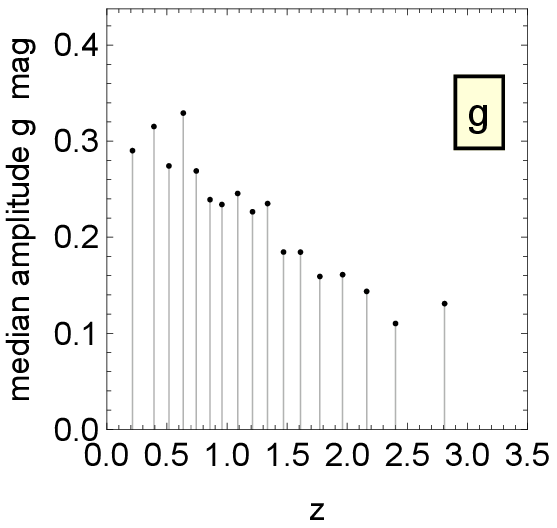}
\includegraphics[width=50mm]{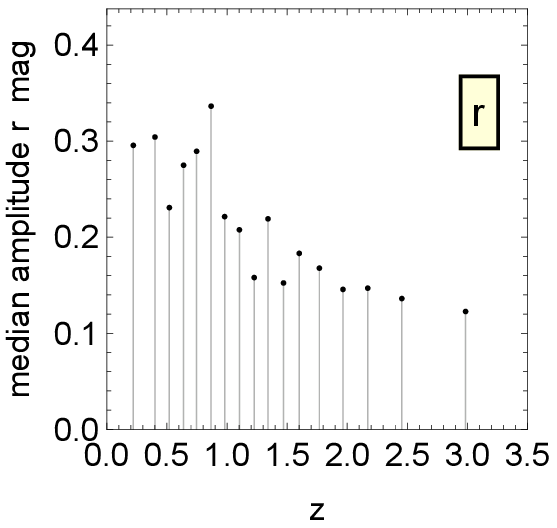}
\includegraphics[width=50mm]{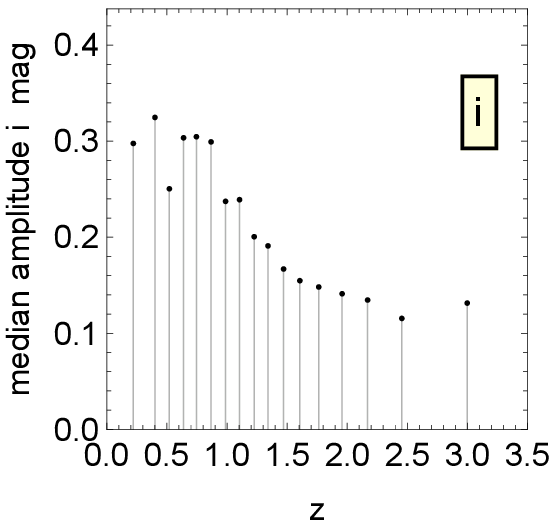}
\includegraphics[width=50mm]{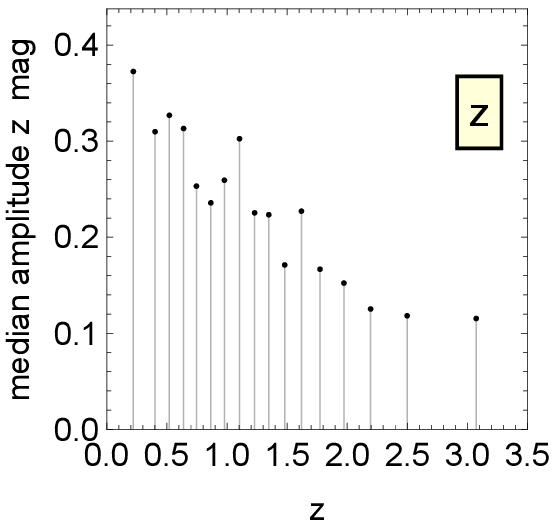}
\includegraphics[width=50mm]{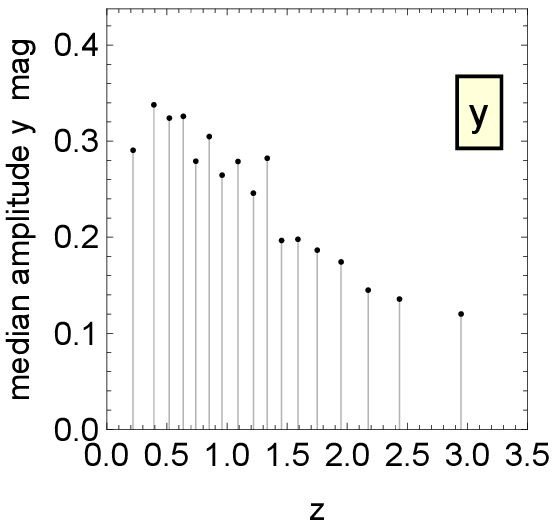}
\caption{Binned median values of variability amplitude $a^{(f)}$ as a function of redshift z
for the five Pan-STARRS bands.}
\label{agz.fig}
\end{figure*}

The intersection of our ICRF3-PS1 sample with the NED catalog includes
2862 objects. Counting objects with nonzero $a^{(f)}$ amplitudes (signifying valid determinations) and with 
observational redshifts $z$, the numbers are 1546, 1694, 1732, 1720, and 1627 for the
$grizy$ bands. The redshifts are of heterogeneous
origin, and here we use them only for large-number statistics estimation. Fig. \ref{agz.fig} shows the binned
median variability amplitude $a^{(f)}$ as a function of redshift. Quasars with redshifts about 0.5 are the most variable
sources, while at redshifts greater than 0.7--1.0, a dramatic drop in the amplitude is observed. This result is strong,
as the same pattern, with small variations, is present in all PS1 bands. The median amplitude of quasars
with redshifts around 3.0, for example, is only one-third of the peak value in the $i$ band. Furthermore, the same
characteristic behavior is seen in higher quantiles of the binned sample, e.g., the 0.75-quantile. As a tentative
explanation of the decline, we offer this consideration. If the power spectrum density of flux variation follows
a monotonously declining power law, most of the variation should be coming from the lowest frequency part
in the range probed by our method, corresponding to a time scale of about 4 years. The time dilation effect shifts
the window of probed frequencies to higher values in the rest frame of the emitter by a factor of $(1+z)$. Thus,
for a high-redshift quasar, we are measuring intrinsically higher-frequency parts of the spectral density distribution.

\begin{figure*}
\includegraphics[width=60mm]{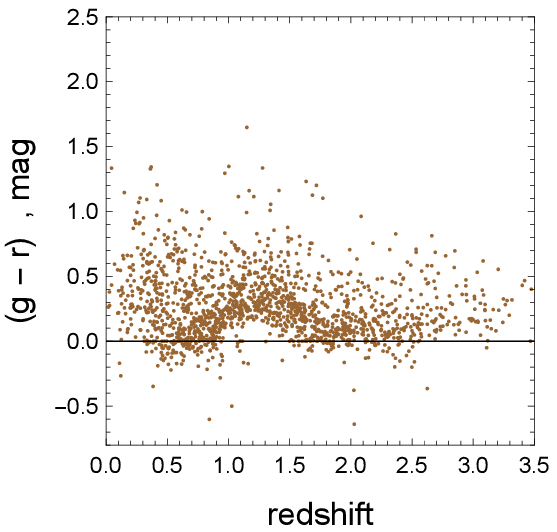}
\includegraphics[width=50mm]{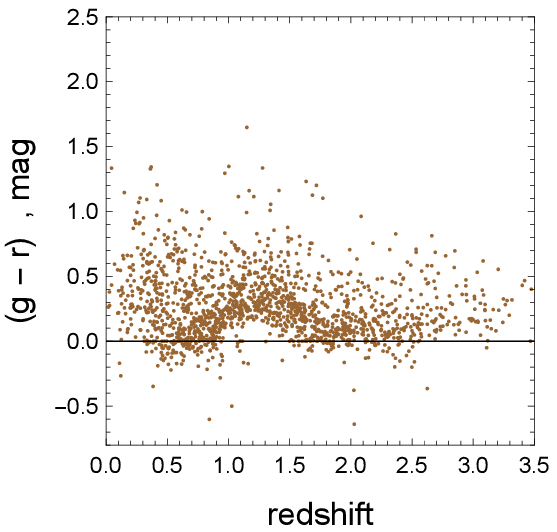}
\includegraphics[width=50mm]{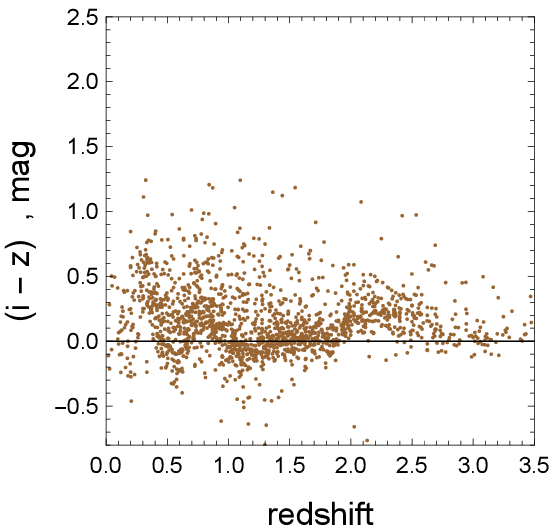}
\includegraphics[width=60mm]{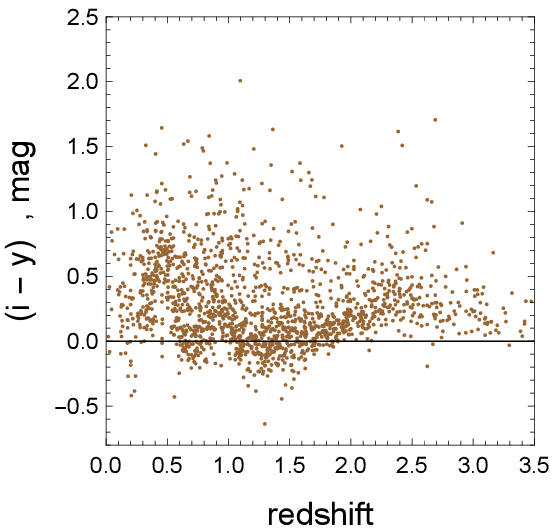}
\caption{Nominal colors of ICRF3 quasars versus redshift.}
\label{col.fig}
\end{figure*}

The homogeneous method of average magnitude estimation allows us to investigate the dependence of observed
colors on redshift. Fig. \ref{col.fig} shows the entire collection of average colors $g-r$, $r-z$, $i-z$,
and $i-y$ for those ICRF3 counterparts that have measured redshifts in the NED. We observe complex, wavy
patterns and distinct correlations, albeit with a significant scatter. Similar patterns in observed color
versus redshift have been seen in other studies \citep{ric, wil}. 
%\emph{\bf References and discussion of emission lines presence in PS1 bands is needed here.}
This is the direct result of bright emission lines and other spectral features redshifted to and out of
the photometric bands. In particular, the blue dip in the $g-r$ relation at $z\simeq0.5$ is caused by the
$\lambda 3000$ bump residing mostly in the $g$ band, but the considerable scatter above the median betrays 
the presence of a significant fraction of intrinsically red objects. The reddening bump at $z\simeq1.2$
is due to the superposition of the Mg II line and the $\lambda3000$ bump in the $r$ band, etc. We note
the close similarity of our results for $g-r$ and $i-z$ to the corresponding relations in 
\citet[][their Fig. 5]{ric}, which comes from the closeness of Pan-STARRS bands to those of the SDSS,
also confirming the high quality of Pan-STARRS photometry and consistency of our method. We also achieve
here a better representation of quasars with $z>2.2$ and extend the coverage to the reddest $y$ band.

\section{Discussion}
\label{dis.sec}
Our main results for a sample of radio-loud ICRF3 quasars with a large range of redshifts indicate that
the statistical properties of these objects are similar to other previously studied sets
of spectroscopically selected objects, most notably, from the SDSS. Previous studies utilizing the Pan-STARRS $3\pi$ survey were mostly focused on refining QSO detection techniques and estimating the PSD slope of large samples \citep[e.g.,][]{morg}, while our study targets specific radio-loud ICRF3 sources. A typical or common variability
amplitude for these radio-loud AGNs is not large at about 0.1 mag. The distribution of amplitudes, however, shows a powerful
tail stretching towards 2 mag and higher, indicating the presence of blazars and other highly
variable kinds. From the photometric point of view, the ICRF3 sample is a complex mixture of various
types, which leaves the possibility to relate the photometric deviants with the observed astrometric
position offsets.

Our information about the relative variability across the Pan-STARRS bands, presented in Table \ref{h.tab}
with three different statistics, appears to be somewhat divergent from the commonly accepted ``brighter-bluer"
phenomenon found in SDSS photometry \citep[e.g.,][]{sch}, in that a convex relation with
increased variability on both ends of the range is found. This may still be veritable because our study
extended to the reddest $z$ and $y$ bands in a consistent way, barely probed before. However, the impact of random observational errors is also greater at the reddest bands, and the increase of variability toward near infrared needs to be verified with better cadenced or more precise measurements.

The strong and confidently detected declining dependence of variability
with redshift (Fig. \ref{agz.fig}) is, perhaps, the most surprising result. The likeliest explanation
is provided by the well-known time dilation effect in combination with a red noise distribution of the
variability power spectrum in rest frame. Indeed, with a limited range of time frequencies and the regularization
weight function in Eq. \ref{wh.eq}, our analysis is most sensitive to variations with periods around 1 -- 4
years in the observer's frame, which correspond to 0.25 -- 1 years in the rest frame for the most distant
quasars in the sample. Their intrinsic variability on these shorter time scales is lower.

While the complicated dependencies of colors on redshift strongly confirm the previous results and their
interpretation (various spectral lines entering and leaving the relevant bands), the knowledge is extended
to the reddest band bordering the near-infrared. Invariably, the empirical relations look sharper at higher
redshifts (Fig. \ref{col.fig}), and the significant one-sided scatter at low $z$ confirms the admixture
of intrinsically red sources in the ICRF3 sample. This is another possible correlation with the
radio-optical position offsets.

\label{lastpage}

\end{document}